\documentclass[12pt]{article}
\let\section=\subsection     \let\subsection=\subsubsection                
\usepackage{graphicx}

\newcommand{\dd}{{\rm d}} 
\newcommand{\bra}[1]{\left\langle #1 \right|}
\newcommand{\ket}[1]{\left| #1 \right\rangle}
\newcommand{\gev}{~{\rm GeV}}
\newcommand{\mev}{~{\rm MeV}}

{\bf }

\begin{document}
\begin{center}
   {\large \bf HIGHER TWIST EFFECTS IN NUCLEI}\\[5mm]
   R.~J.~FRIES \\[5mm]
   {\small \it  Department of Physics, Duke University \\
   Box 90305, Durham, NC 27708, USA \\[5mm] }
   {\small \it Institute for Theoretical Physics, University of Regensburg, \\
   93040 Regensburg, Germany\\[8mm] }
\end{center}

\begin{abstract}\noindent
  This talk serves as an introduction to higher twist effects in
  nuclei. We want to discuss how perturbative QCD can be applied to processes
  involving heavy nuclei by taking into account multiple scattering.
\end{abstract}

\section{Introduction}
Perturbative Quantum Chromodynamics (pQCD) has been established as an 
extremely successful theory to describe phenomena related to scattering 
reactions off hadrons at large momentum transfer. pQCD calculations combined 
with the available parameterizations of parton distributions enable us to
explain a large set of data including those from deep inelastic lepton nucleon
scattering (DIS)
$l + N \longrightarrow l + X$ and the famous Drell Yan process (DY) $N + N 
\longrightarrow l^+ + l^- + X$.

pQCD is in fact a rather strict theory in the sense that no assumptions arising
from any model enter.
There are quantities of non-perturbative nature which cannot be described by 
perturbation theory, e.g.\ the bound states of QCD. 
Nevertheless there exists a quite rigorous way to separate perturbative (short
range) and non-perturbative (long range) physics in a scattering reaction in a
proper way.
Let us consider DIS as an example. A factorization theorem \cite{CSS:mueller} 
enables us to shift all non-perturbative physics into a set of well-defined, 
gauge-invariant (i.e.\ observable) and universal (i.e.\ process independent) 
quantities. These quantities can be expressed by matrix elements of parton
operators between hadron states.

In principle an infinite number of such matrix elements could enter our 
calculation and spoil the usability of the factorization theorem. 
But we can establish a hierarchy between the matrix elements in terms of an
expansion in inverse powers of the momentum transfer. In more detail the 
quantity we expand in is $\lambda / Q$ where $Q$ is the perturbative hard 
scale, e.g.\ the virtuality of the virtual photon in DIS, and 
$\lambda$ (for massless QCD) has to be some non-perturbative (hadronic) scale.
The perturbative scale $Q$ makes the coupling running and has to be large (at
least a few GeV) in  order to make perturbation theory applicable while 
$\lambda$ is of the order of $\Lambda_{QCD}\approx 200\mev$. Under this
circumstances $\lambda / Q \ll 1$ and this expansion, called the \emph{twist} 
expansion, will do well. This is the definition of twist we will use here.

The factorization theorem tells us
that the left diagram in Fig.~1 gives the leading contribution to DIS in the
twist expansion (and also in the $\alpha_s$ expansion).\footnote{We are not
going to discuss the subtleties arising from the fact that we have two
expansions, one in $\lambda/Q$ and one in $\alpha_s \sim \ln \lambda/Q$, which 
are obviously not independent.}
\begin{center}
   \includegraphics[height=2.6cm]{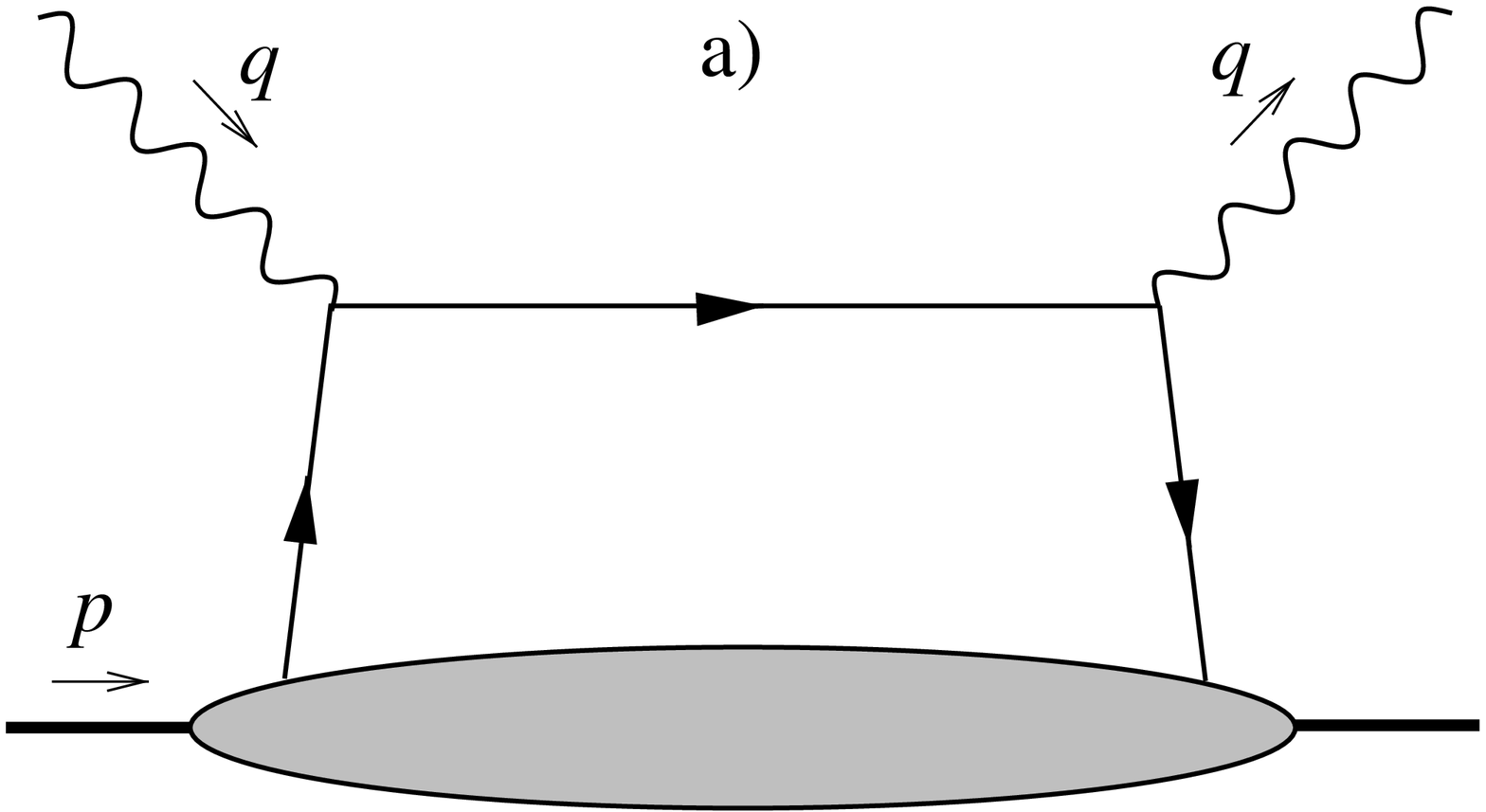} \hspace{3em}
   \includegraphics[width=3.5cm]{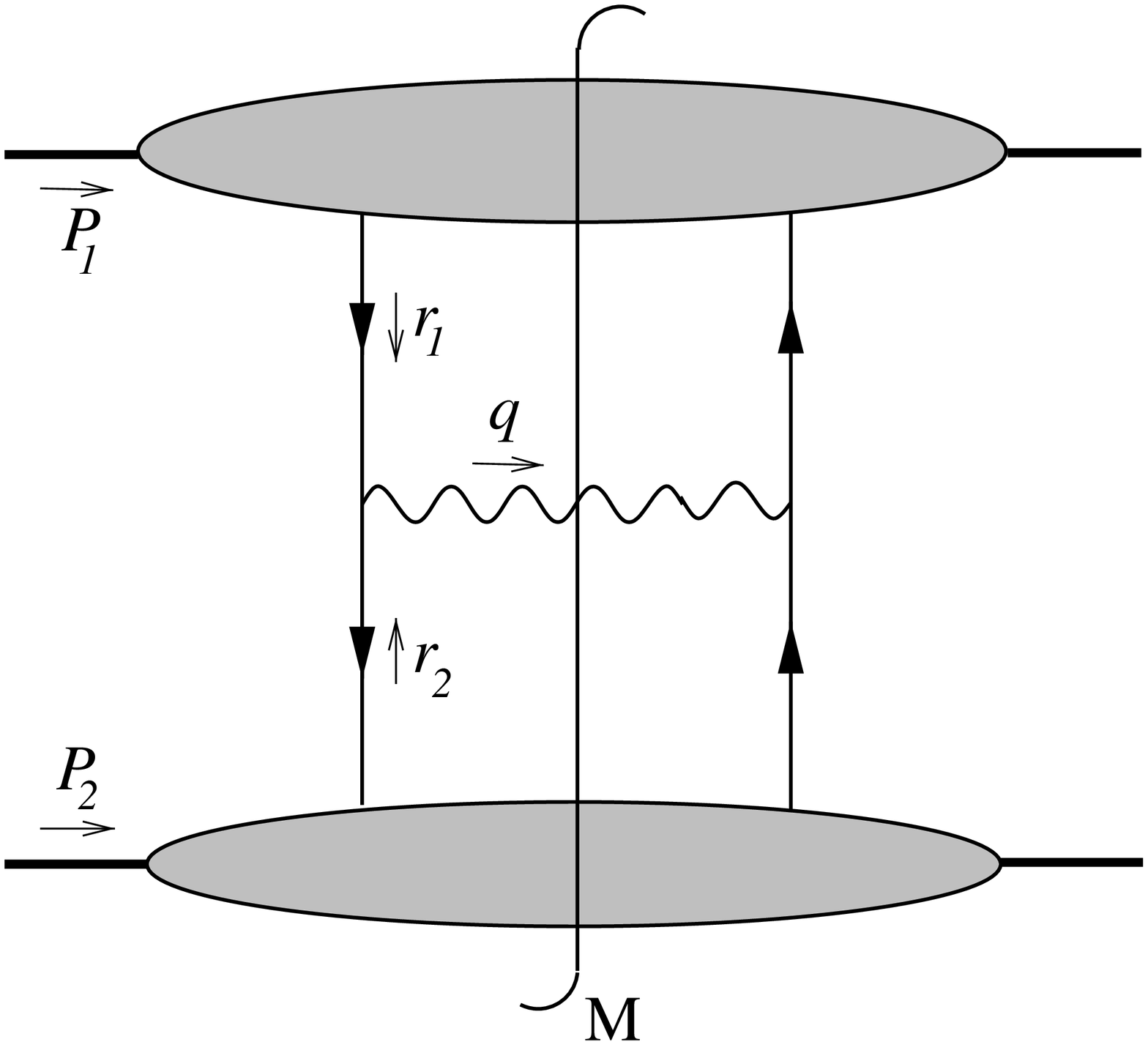}\\[0.5cm]
   \parbox{14cm}
        {\footnotesize 
        Fig.~1: Left: leading twist contribution to deep inelastic lepton 
        nucleon scattering. The virtual photon
        emitted from the lepton (not shown) scatters off one quark inside the 
        nucleon. 
        Right: leading twist contribution to Drell Yan pair production in
        nucleon nucleon collisions. Each nucleon emits a quark or antiquark
        which annihilate into a virtual photon, which then
        decays into a lepton pair (not shown).}
\end{center}

Note that the leading twist contribution always consists of one hard 
scattering on the parton level, here the scattering of the photon off one 
quark from the nucleon. The non-perturbative part is described by a matrix 
element which encodes the process of taking one quark out of the nucleon and 
putting it back (in the complex conjugated graph). It is given by
\begin{equation}
  f_{q} (\xi) =
  \int \frac{\dd y^-}{2\pi} e^{i \xi P^+ y^-} \,
  \frac{1}{2}\bra{P} \bar{q}(0) \gamma^+ q(y^-) \ket{P},
\end{equation}
where $q$ are quark field operators and $P$ is the momentum of the nucleon.
The Bjorken variable $\xi$ gives the momentum fraction of the nucleon 
that is carried by the quark in an infinite momentum frame. $f_{q} (\xi)$
is then just the famous quark distribution function in the nucleon. It has
a parton model interpretation as the probability to find a quark with momentum
fraction $\xi$ in the nucleon. 
This leading twist contribution to DIS is also called twist-2. The next order 
(next-to-leading twist) would involve matrix elements of twist-4, i.e.\ ${\cal
O}(\lambda^2/Q^2) $.\footnote{
Twist-3 vanishes for unpolarized scattering.} The cross section for DIS can
then be written as a convolution of a parton distribution $f_a$ with a parton
cross section $\sigma_{l+a}$ --- which itself is a series in $\alpha_s$
--- plus power corrections which arise from matrix elements which differ from
parton distributions.
\begin{equation}
  \sigma_{l+p} = \sum_a f_a \otimes \sigma_{l+a} + 
   {\cal O} \left( \frac{\lambda^2}{Q^2} \right) 
\end{equation}

For the Drell Yan process also a factorization theorem holds which gives
the leading twist contribution as a simple annihilation of quarks and 
antiquarks to produce a virtual photon, see Fig.~1 (right).
Note again that leading twist (twist-2) involves only one hard scattering 
on the parton level (the annihilation), while two matrix elements, one for
each nucleon, are in the game. These matrix elements define exactly the same
parton distributions as in DIS. This powerful property is called universality.
Measuring a parton distribution in one process gives us predictive power for 
all other processes where this parton distribution enters.

\section{Multiple Scattering and Nuclear Enhancement}

What changes if we replace single hadrons by nuclei and look at $e+A$, $p+A$ 
or $A+A$ collisions? 
The factorization theorems should hold also here, but for many observables
the picture of one hard scattering without rescattering on the 
parton level seems not to be the dominant one. E.g.\ for Drell Yan we expect 
initial state interactions when the quark and the antiquark participating in 
the annihilation have to traverse a large piece of nuclear matter. 
In the framework of pQCD multiple scatterings are exactly \emph{higher twist}
corrections to the single scattering (leading twist) process.

At a given scale $Q^2$ the power corrections by 
higher twist effects are enhanced because of the nuclear size. This 
enhancement cannot stem from the hard (short range) part of the cross section,
but must arise from the non-perturbative (long range) matrix elements. We will
demonstrate later how the matrix elements can test the nuclear size.

It was an idea mainly advocated by Luo, Qiu and Sterman \cite{LQS} that, 
e.g.\ on the twist-4 level, there are matrix elements which scale like 
$\lambda^2 A^{1/3}/ Q^2$ where $A$ is the mass number of the nucleus, whereas 
there are others which are not sensitive to the nuclear size and behave just
like $\lambda^2 /Q^2$ as usual. For large nuclei we can assume
that $A^{1/3} \gg 1$ and conclude that only those higher twist corrections,
which show an additional scaling with the nuclear size, called \emph{nuclear 
enhanced}, are important and we safely omit all others.
We rearrange our twist series in such a way, that we have an expansion in
powers of $\lambda^2 A^{1/3} /Q^2$, i.e.\ for each power of $Q^{-2}$ we have an
additional power of $A^{1/3}$ (maximal nuclear enhancement). Terms with less 
powers in $A^{1/3}$ than in $Q^{-2}$ are subleading.

Twist corrections normally are a tricky business and very difficult to handle.
Starting from the parton distribution in Eq.~(1) we can e.g.\ generate higher 
twist matrix elements just by inserting covariant derivatives acting on the 
parton fields. Unfortunately the number of matrix elements contributing at a 
certain level of twist can be large and in addition most of these matrix 
elements have no probabilistic interpretation in a parton picture.

This changes dramatically if we demand that a matrix element should be
sensible to the size of an extended medium. Extended here means that the
extension $L$ should be much larger than the confinement radius $R_0$,
$L\gg R_0$, what is true for large nuclei. One can show that only
matrix elements with additional pairs of parton operators can contribute.
For twist-4 schematically we have e.g.\ $\bra{P} (\bar q q) (FF)\ket{P}$ or
$\bra{P} (FF) (FF)\ket{P}$, where color, spinor and vector indices for each
pair are contracted in the same way as in the parton distribution 
respectively. $F$ here is a gluon field strength. Therefore these matrix 
elements are limited in number and have a
straight forward interpretation as correlators between partons.

In Fig.~2 we give an example for single and double scattering in the DY 
process at large transverse momentum. Here the quark (antiquark) from the
single hadron has the opportunity to scatter off an additional gluon from the
nucleus before annihilating with an antiquark (quark).
\begin{center}
   \includegraphics[width=5cm]{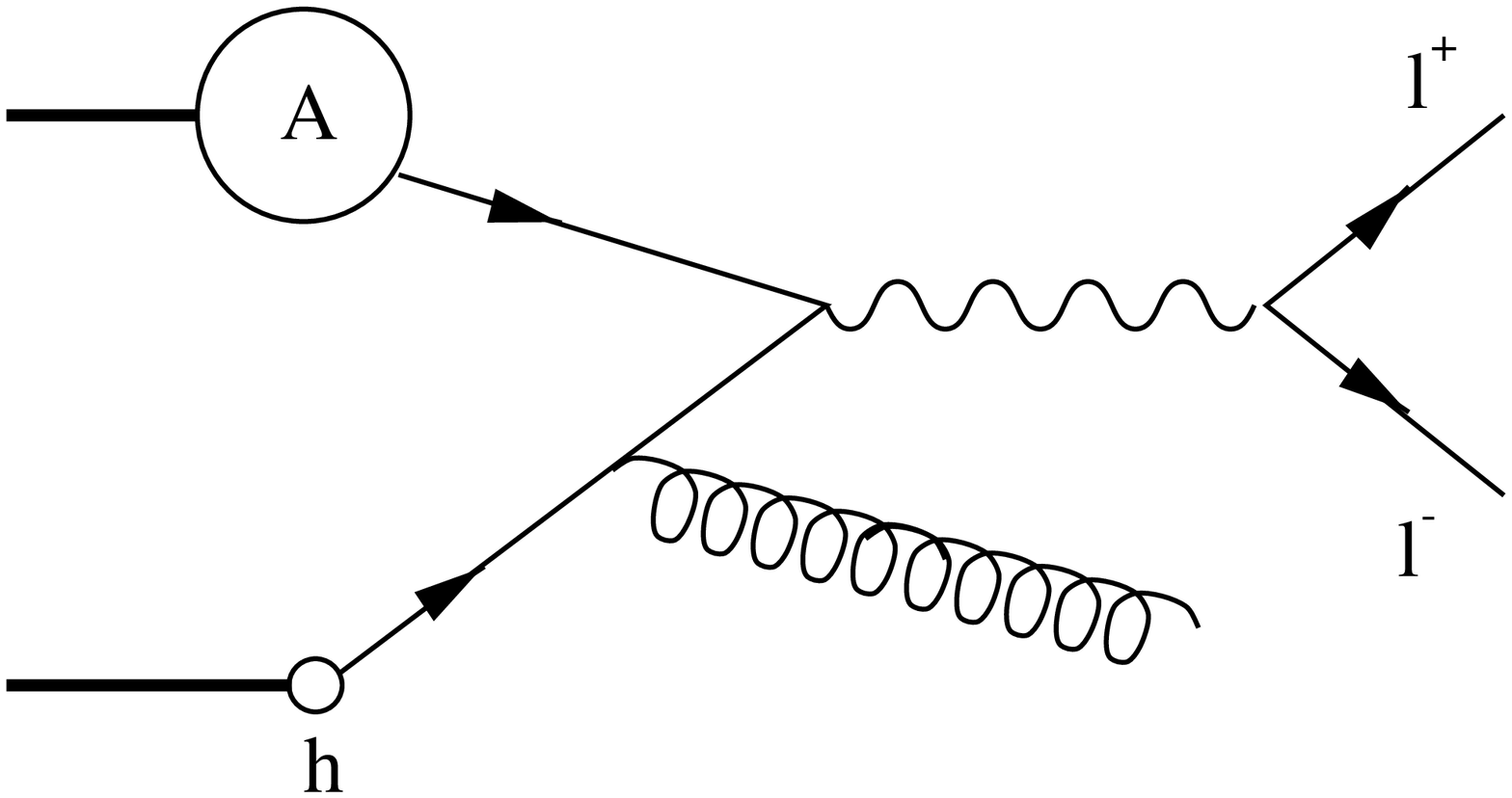} \hspace{3em}
   \includegraphics[width=5cm]{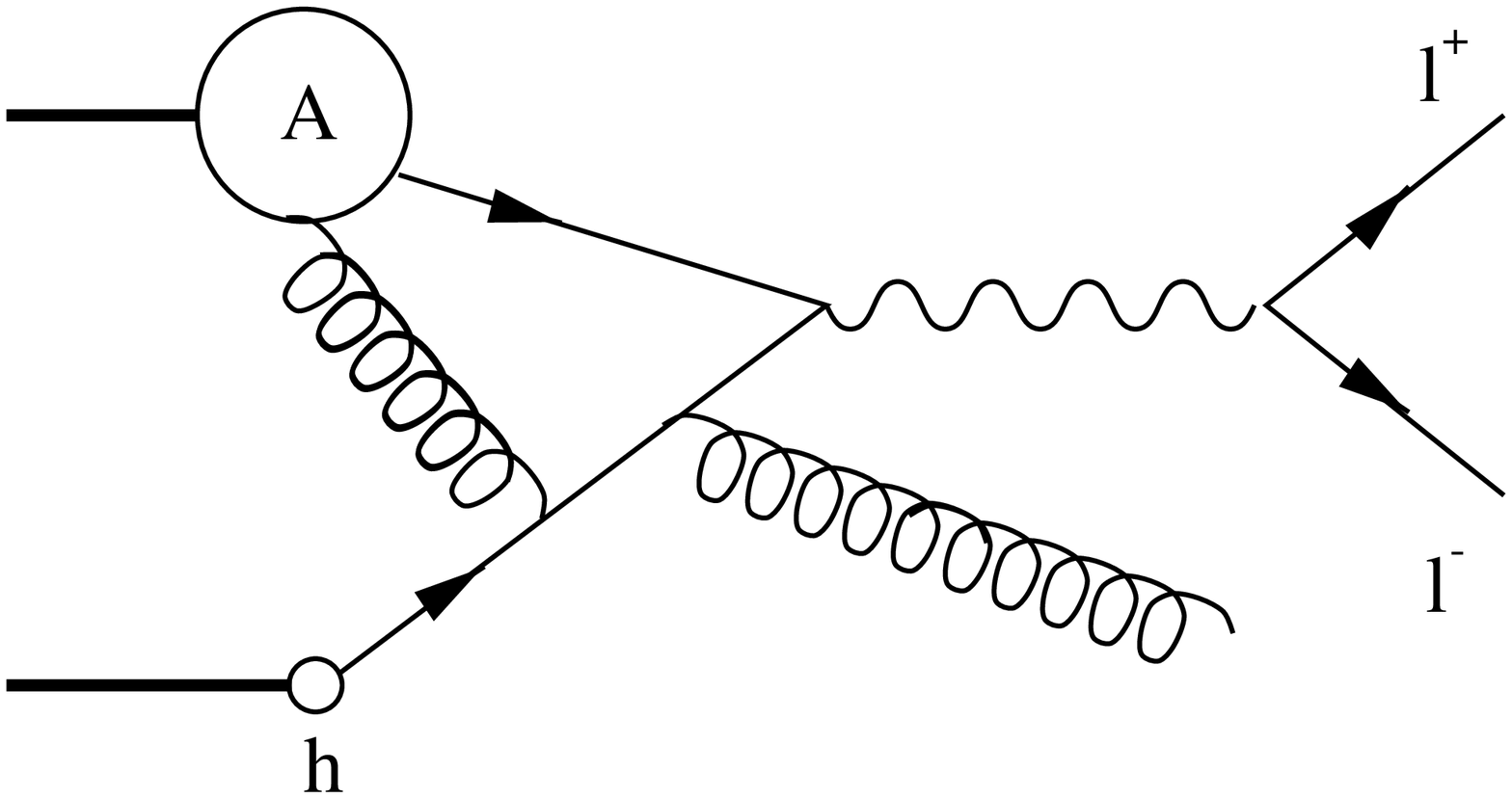}
   \\[0.5cm]
   \parbox{14cm}
        {\footnotesize 
        Fig.~2: Possible diagrams for single scattering (left) and double 
        scattering (right) contributing to DY pair production in $p+A$ 
        collisions at high transverse momentum. Note that both diagrams are
        next-to-leading order in $\alpha_s$ respectively, since we radiate 
        an extra gluon in order to generate finite transverse momentum.}
\end{center}

The onset of nuclear enhancement can be seen in $p+A$ experiments. 
The best known example is the famous Cronin effect \cite{Cron:75,
E772:90dy}, which is a nuclear enhancement of the production cross section
of Drell Yan pairs, $J/\psi$s or other particles at large transverse momentum,
whereas at low transverse momentum enhancement id missing. 
The enhancement is a clear sign of multiple scattering. Another sign is the 
transverse momentum broadening of jets or Drell Yan pairs. Fermilab data
indicate a rise of the ratio of second and first moment of the transverse 
momentum spectrum which is compatible with an $A^{1/3}$ behaviour 
\cite{McGMP:99}.

Let us emphasize that single scattering contributions in principle scale
with $A$, i.e.\ just they scale with the volume the reaction can take place.
The nuclear enhanced double scattering picks up an 
\emph{additional} power of the nuclear radius and therefore scales like 
$A^{4/3}$ and so on. With this mind one would expect an overall enhancement
of all cross sections but experimentally this is clearly not the case. 
Indeed most cross
sections are not enhanced like e.g.\ DY at low transverse momenta. If pQCD 
intends to explain the enhancement by the introduction of new matrix elements, it
should also account for the absence of enhancement in other observables.
Again this works fine. We will see that \emph{interference} seems to be very
important. This can spoil the enhancement in certain observables.

\section{An example: double scattering in Drell Yan}

Let us discuss the Drell Yan process in $p+A$ as an example. We would like
to calculate the effect of double scattering at large transverse momentum 
\cite{Guo:98ht,FSSM:99,FSSM:00}.
The DY cross section with full kinematic dependence on Mass $Q$, transverse
momentum $q_\perp$ and rapidity $y$ of the lepton pair and on the angular 
distribution of the lepton pair can be decomposed into four structures
parametrized by four helicity amplitudes $W_{TL}$, $W_L$, $W_\Delta$ and
$W_{\Delta\Delta}$.
\begin{eqnarray}
  \frac{\dd \sigma}{\dd Q^2 \,\dd q_\perp^2 \,\dd y \,\dd\Omega} &=&
  \frac{\alpha^2_{em}}{64 \pi^3 S Q^2}
  \Big( {W_{TL}}\left(1+\cos^2\theta\right) +
     {W_L}\left(\frac{1}{2}-\frac{3}{2}\cos^2\theta\right) + \nonumber \\
     & & {W_\Delta}\left(\sin 2\theta\cos\phi \right) +
     {W_{\Delta\Delta}}\left(\sin^2\theta\cos 2\phi \right) \Big)
\end{eqnarray}
If we integrate over the angles, only the term proportional to $W_{TL}$ 
contributes to the cross section. Fig.~3 shows two typical examples for
graphs for twist-4. At this order of twist we not only have to take into 
account double scattering but also interference of single and triple 
scattering (which we will include into the term double scattering in the 
following since the matrix elements are the same).
\begin{center}
   \includegraphics[width=5cm]{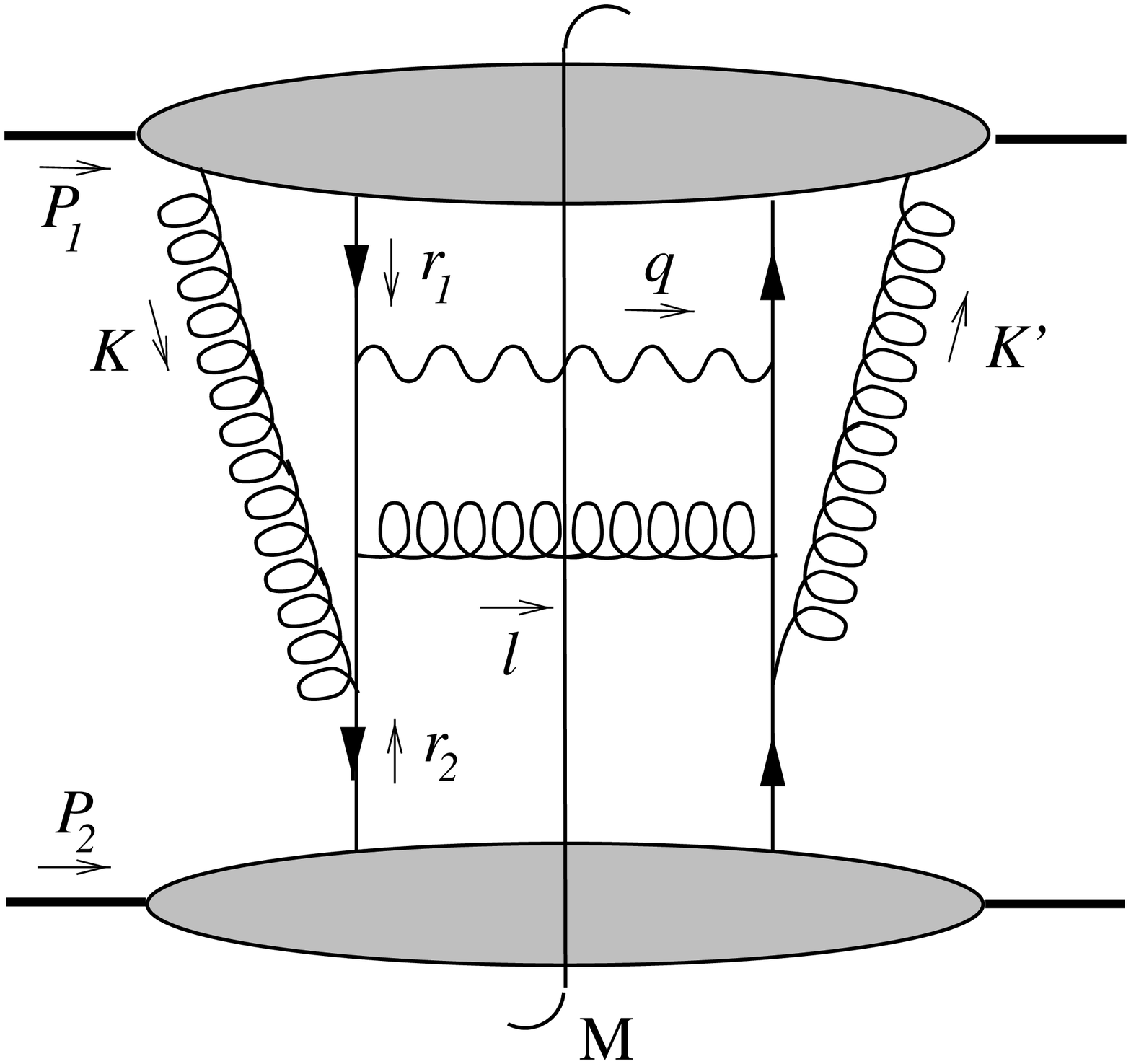} \hspace{3em}
   \includegraphics[width=5cm]{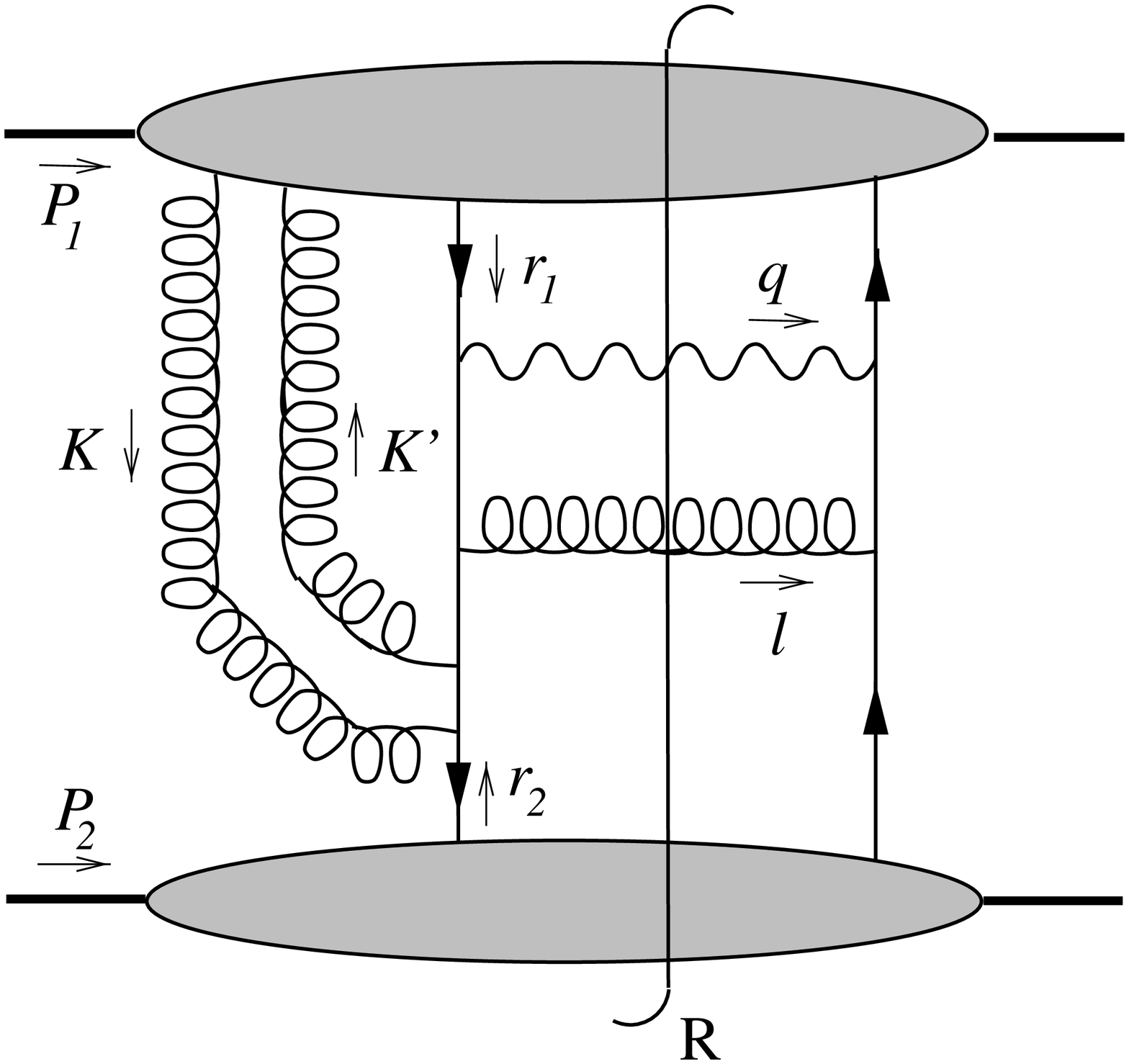}
   \\[0.5cm]
   \parbox{14cm}
        {\footnotesize 
        Fig.~3: Two diagrams contributing to twist-4 at large transverse 
        momentum. The left diagram shows an ordinary double scattering 
        with a quark gluon pair from the nucleus on the left and right 
        (complex conjugated) part of the diagram. The right diagram shows
        a possible interference between triple scattering (two gluons and one
        quark on the left part) and single scattering (one quark on the right
        side).}
\end{center}

It turns out that there are two different contributions to double scattering.
Both partons from the nucleus can be hard (double hard scattering) or one can
be hard and one soft (soft hard scattering). The final formulae for 
the helicity amplitudes $W_i$ including the partons $a$ and $b$ from the 
nucleus and parton $c$ from the single nucleon are
\begin{equation}
  W^{DH,ab+c}_i =
  \int_B^1 \frac{\dd\xi_2}{\xi_2} 
  {{f_{c}(\xi_2)}}{{ T^{DH}_{ab}(x_a,x_h)}}
  {{H^{DH,ab+c}_{i}(x_a,x_h,\xi_2)}},
\end{equation}
\begin{eqnarray}
  W_{i}^{SH,ab+c} &=&  
 \int_B^1 \frac{\dd \xi_2}{\xi_2} 
   {{f_{c} (\xi_2)}}  \nonumber \\
  &&\big( -\frac{g^{\lambda\kappa}}{2} \big) 
  \frac{\dd^2}{\dd K_\perp^\lambda \dd K_\perp^\kappa}\Big\vert_{K_\perp=0} 
  {{T^{SH}_{ab}(x_b)}} {{H^{SH,ab+c}_{i}(x_b,x_s,\xi_2)}}
\end{eqnarray}
for double hard and soft hard scattering respectively.
$f_c$ are the usual parton distributions for the nucleon, $H^{ab+c}$ is
the cross section on the parton level (the hard part of the cross section)
and the $T_{ab}$ are the new matrix elements describing two partons in the 
nucleus.

Let us shortly discuss where the distinction between soft hard and double 
hard arises from. We have to integrate over longitudinal momenta of parton 
lines connecting soft and hard parts of the diagram and we have propagators 
in the hard part which can provide poles in the integrands. An example 
is given in Fig.~4. Both propagators marked by circles give poles for the 
integration of the gluon momentum fraction $x$.
\begin{center}
   \includegraphics[width=5cm]{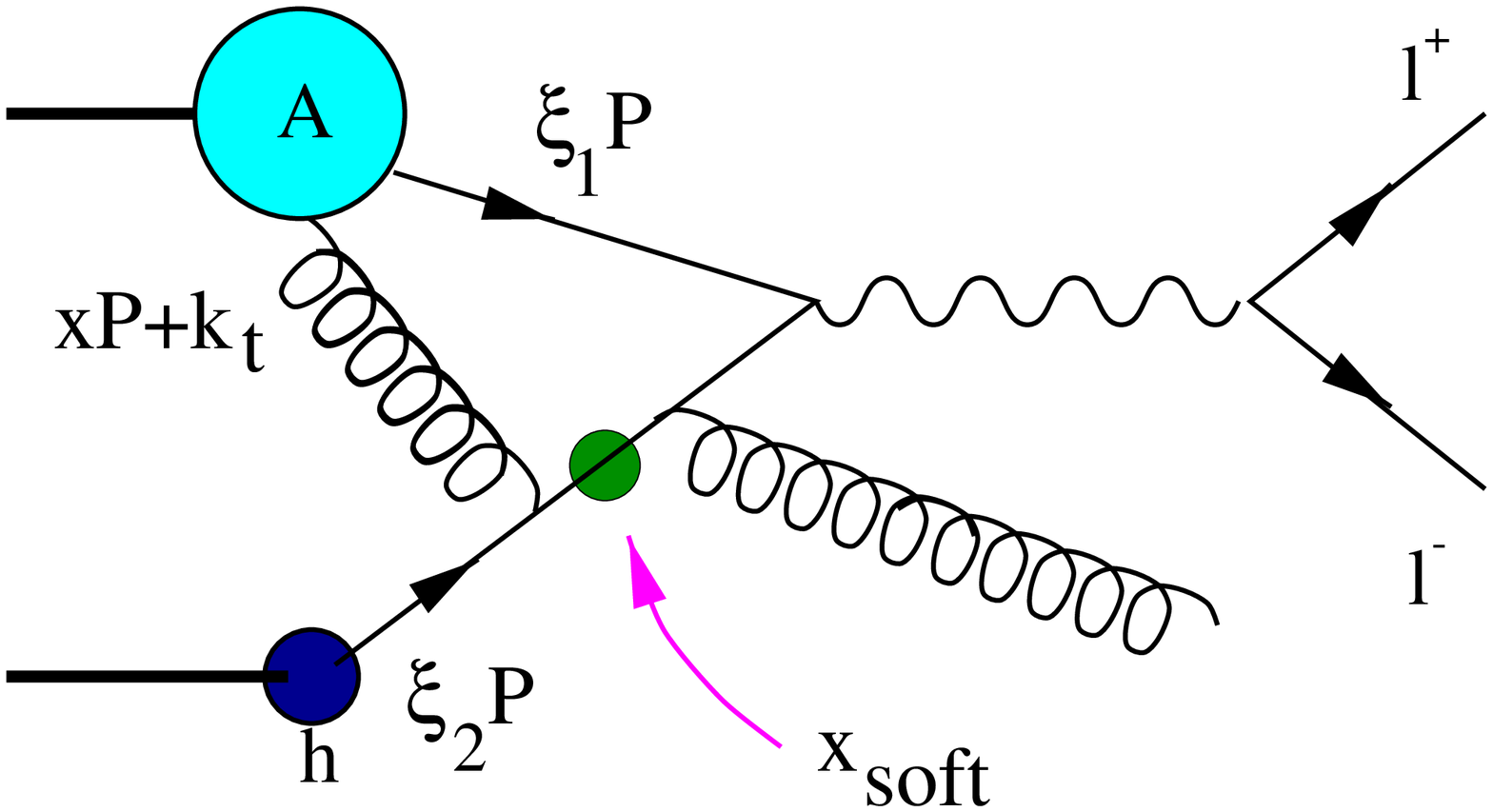} \hspace{3em}
   \includegraphics[width=5cm]{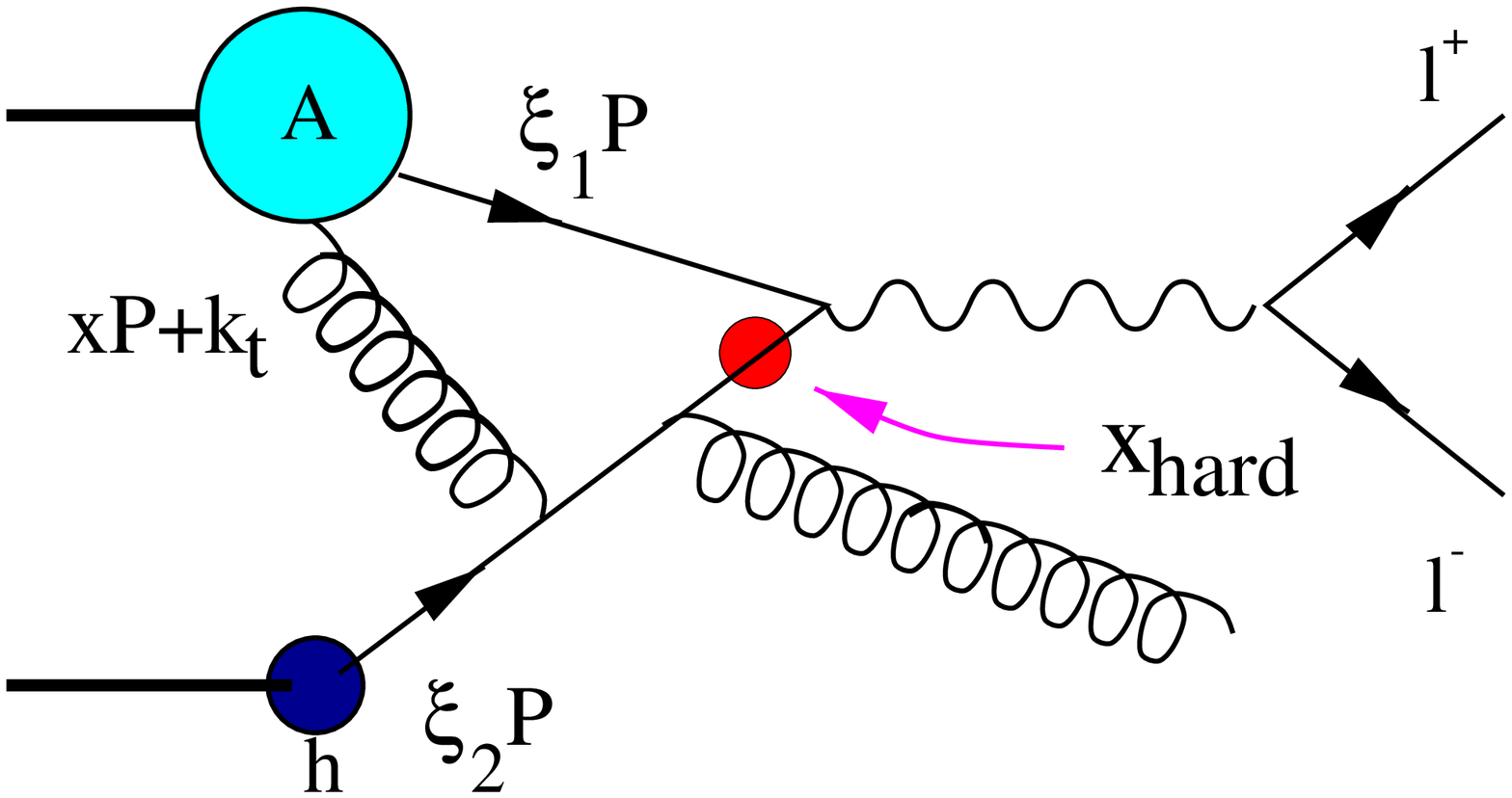}
   \\[0.5cm]
   \parbox{14cm}
        {\centerline
        {\footnotesize 
        Fig.~4: Soft pole (left) and hard pole (right) indicated by circles.}}
\end{center}
The upper pole fixes $x$ to a large (finite) value $x_{hard}$, whereas the 
lower pole gives $x_{soft}\sim K^2_\perp / S \sim 0$ where $K_\perp$ is the 
intrinsic transverse momentum of the gluon. Carrying out the pole integration 
gives a sum of both residues:
\begin{eqnarray}
  M &\sim& \int \dd x \left(\frac{1}{x-x_{{soft}}+ i \epsilon} \right)
                  \left(\frac{1}{x-x_{{hard}}+ i \epsilon} \right)
  F(x, k_t, \xi_1) \nonumber \\
  &=& \left(\frac{F(x_{{soft}}, k_t, x_{tot} - x_{{soft}})
  }{x_{{soft}} -x_{{hard}}} \right) - 
  \left(\frac{F(x_{{hard}}, k_t, x_{tot} - 
  x_{{hard}})}{x_{{soft}}-x_{{hard}}} \right) \\                 
  &=& M_{{{soft}},hard} - M_{{{hard}},hard} \nonumber
\end{eqnarray}
Note the minus sign between both terms. To get the cross section we have to 
square the term above. The hierarchy $x_{{hard}} \gg x_{soft}$
can only be ensured as long as the transverse momentum $q_\perp$ is large.
In this case the interference between both residues is negligible and the
cross section is given just by the sum of squares 
$M^2_{{{soft}},hard} + M^2_{{{hard}},hard}$. The first term is the soft hard
contribution (soft gluon) the second term is the double hard contribution
(hard gluon).

However if $q^2_\perp \ll Q^2$ then $x_{hard}$ starts to approach $x_{soft}$ 
and both residues start to cancel each other. This is the explanation of the 
important statement already given above:
for small transverse momenta the double scattering contribution undergoes
destructive interference and there is no nuclear enhancement from double
scattering in that kinematic region.

Let us now discuss the twist-4 matrix elements. The double hard 
matrix elements depend on the momentum fractions $\xi$ and $x$ of both hard 
partons. The quark gluon correlator e.g.\ is
\begin{eqnarray}
  T^{DH}_{qg}(\xi,x) &=& { 
  \frac{1}{x} \int \dd z_4^-
  \frac{\dd z_3^-}{2\pi} \frac{\dd z_1^-}{2\pi} }\Theta(z_1^- -z_3^-)
  \Theta(-z_4^-) \\
  && e^{i\xi P_1^+ z_1^-}   
  e^{ix P_1^+(z_3^--z_4^-)} {\frac{1}{2}}
  \bra{P_1} F^{\omega +}(z_4^-) F^{+}_{\quad\omega}(z_3^-) 
  \bar{q}(0) \gamma^+  q(z_1^-)  \ket{P_1}.  \nonumber
\end{eqnarray}  
The same correlator for soft hard scattering reads
\begin{eqnarray}
  T^{SH}_{qg}(\xi) &=& \int \dd z_4^-
  \frac{\dd z_3^-}{2\pi} \frac{\dd z_1^-}{2\pi} \Theta(z_1^- -z_3^-)
  \Theta(-z_4^-) 
  \\ 
  && e^{i\xi P_1^+ z_1^-}   
  \frac{1}{2}\bra{P_1} F^{\omega +}(z_4^-) F^{+}_{\quad\omega}(z_3^-) 
   \bar{q}(0) \gamma^+  q(z_1^-)  \ket{P_1}. \nonumber
\end{eqnarray}
It depends only on the momentum of the hard parton. The $\Theta$-functions
ensure causality. 
PQCD is not able to predict these matrix elements from first principles and
they are not measured up to now. In order to make any numerical statements
about the size of the nuclear enhanced corrections we have to rely on models
at this stage. However the important point is, we can extract the scaling with
the nuclear size. To do this we have to take into account the colour structure
of the operators, the oscillating exponential factors and the boundaries of the
integrals. An analysis gives that both matrix elements above have one free
integration which can test the extension of the nucleus.
Pictorially this is the distance between the two parton pairs
$\bar q q \> \leftrightarrow \> FF$.
From that we assume that the matrix elements can be modelled by ordinary 
parton distributions by setting
\begin{eqnarray}
  T^{DH}_{ab}(\xi,x) &=& C A^{4/3} f_a(\xi) f_b(x), \\
  T^{SH}_{ab}(\xi) &=& \lambda^2 A^{4/3} f_a(\xi).
\end{eqnarray}
$C$ and $\lambda^2$ are normalization constants of the order of 
$\Lambda_{QCD}$.

At the moment our goal must be to gather experimental information about the
new matrix elements and therefore our study for RHIC \cite{FSSM:99,FSSM:00}
intends to look for observables where this can be done.
Independent of the models which one plugs in for the matrix elements one can 
establish some results:
Double hard scattering has trivial angular dependence like the leading twist,
leading $\alpha_S$ process. It seems to confirm the picture of two independent
binary collisions, first $q+g \rightarrow q+g$ and then 
$q+\bar q \longrightarrow \gamma^*$. Confer the right diagram of Fig.~4: the
pole splits the diagram into two independent subprocesses. Also the so called
Lam Tung sum rule $2W_{\Delta\Delta}=W_L$, a long standing leading twist 
prediction \cite{LT:80}, is respected by double hard scattering.
On the other hand the results on soft hard scattering are more complicated and
violate the Lam Tung relation. Another important point is that the leading 
twist result for $W_{\Delta}$ is nearly zero (it vanishes for $p+p$ from 
symmetry reasons), while there is a non vanishing twist-4 contribution. 
These are the most important model independent results which could allow 
a glance at the new matrix elements in forthcoming data. At the end of this
section we show two results for proton gold collisions at RHIC. In Fig.~5 
the rapidity distributions of the amplitudes $W_{TL}$ and $W_{\Delta}$ are
given. Note that at RHIC energies soft hard seems to be dominated by the double
hard process.
\begin{center}
   \includegraphics[width=5cm]{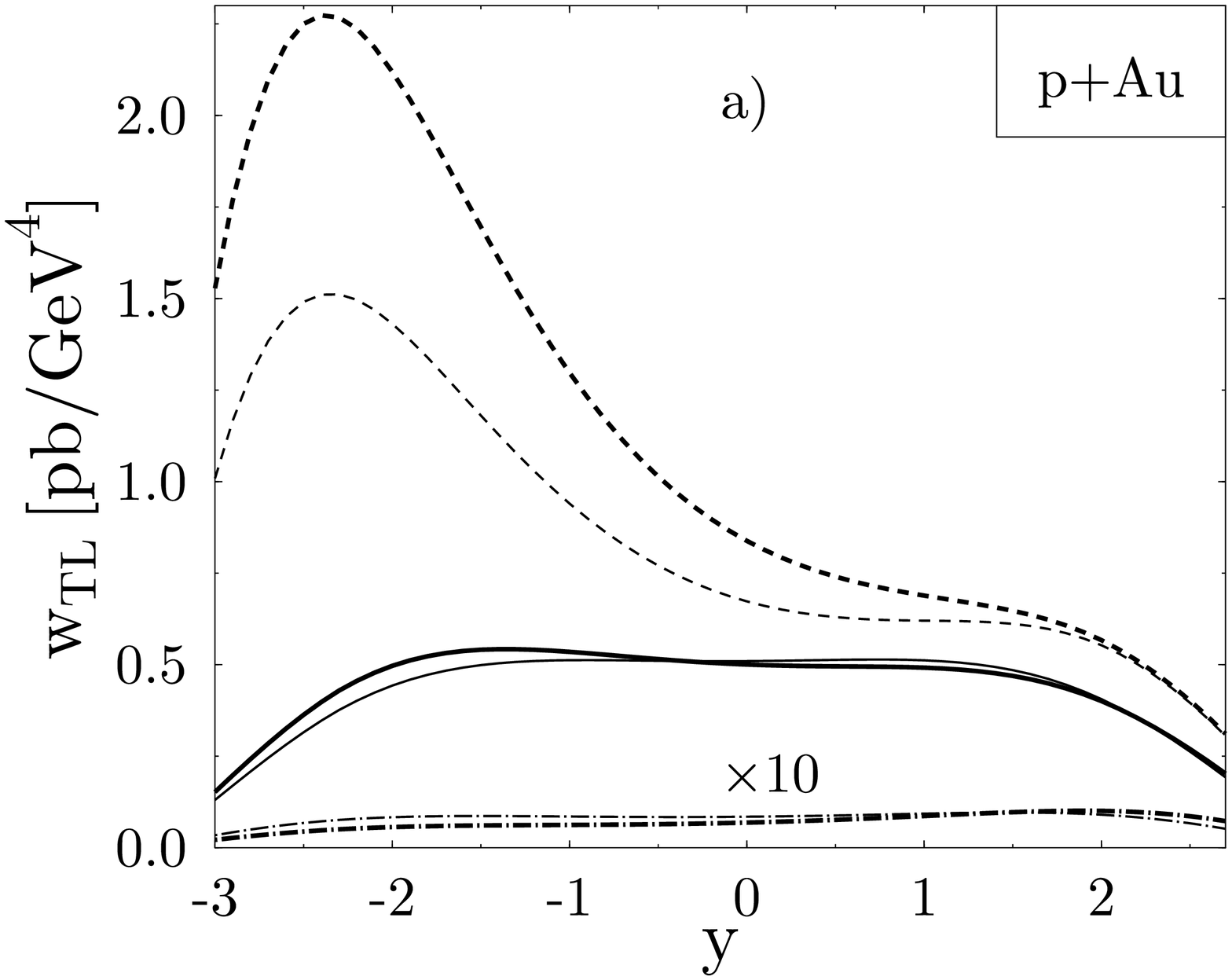} \hspace{3em}
   \includegraphics[width=5cm]{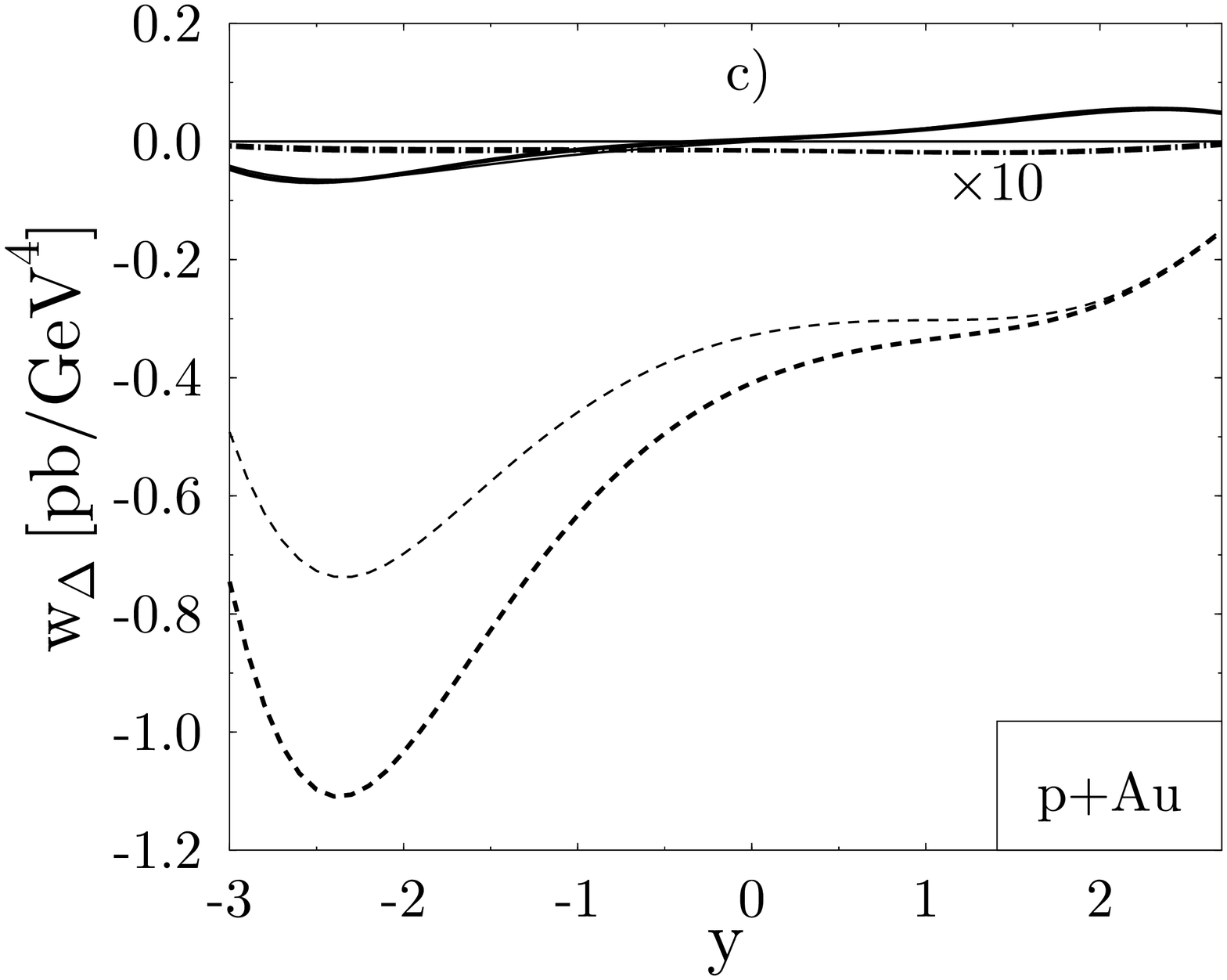}
   \\[0.5cm]
   \parbox{14cm}
        {\footnotesize 
        Fig.~5: Results for $100\gev\>Au + 250\gev\>p$ collisions at RHIC.
        Cross sections for helicity channels $W_{TL}$ (left) and $W_{\Delta}$
        (right) as a function of rapidity $y$ at $Q=5\gev$ and $q_{\perp}=
        4\gev$. Single scattering (solid lines), double hard (dashed) and
        soft hard (dot dashed) for two different choices of parton 
        distribution parameterizations (CTEQ3M and CTEQ3M+EKS98) are shown.}
\end{center}

\section{Further topics}

The transverse momentum broadening of DY pairs was also explained recently
by X.~Guo within this formalism \cite{Guo:98jb}. For this observable one is
sensible to small transverse momenta and one has to calculate graphs like
in Fig.\ 6 (left)
with no additional parton in the final state. The ratio of second and first
moment of the transverse momentum spectrum is surprisingly easy
\begin{equation}
  \frac{\langle q_\perp^2 \rangle}{\langle q_\perp^0 \rangle} =
  \frac{4\pi^2 \alpha_s}{3} \frac{f_c \otimes 
  T^{SH}_{a g}}{f_c \otimes f_{a}} = \frac{4\pi^2 \alpha_s}{3} A^{1/3}
  \lambda^2
\end{equation}
with partons $a$ and $c$ being quark and antiquark or vice versa. The last 
equal sign holds for our model for the soft hard matrix elements. One may note 
that the pQCD result indeed rises with $A^{1/3}$ compatible with the data.

In the low $q_\perp$ regime one would like to sum the effect of multiple
scattering which would be a resummation of higher twist. Diagrams like
in Fig.~6 (right) introduce matrix elements which are correlators with an 
arbitrary number $n$ of gluon fields $\langle P|(\bar q q) (FF)^n |P\rangle$. 
If we extrapolate the model above we can establish connections between matrix 
elements of different $n$ and are able to carry out the summation.
\begin{center}
   \includegraphics[height=2.8cm]{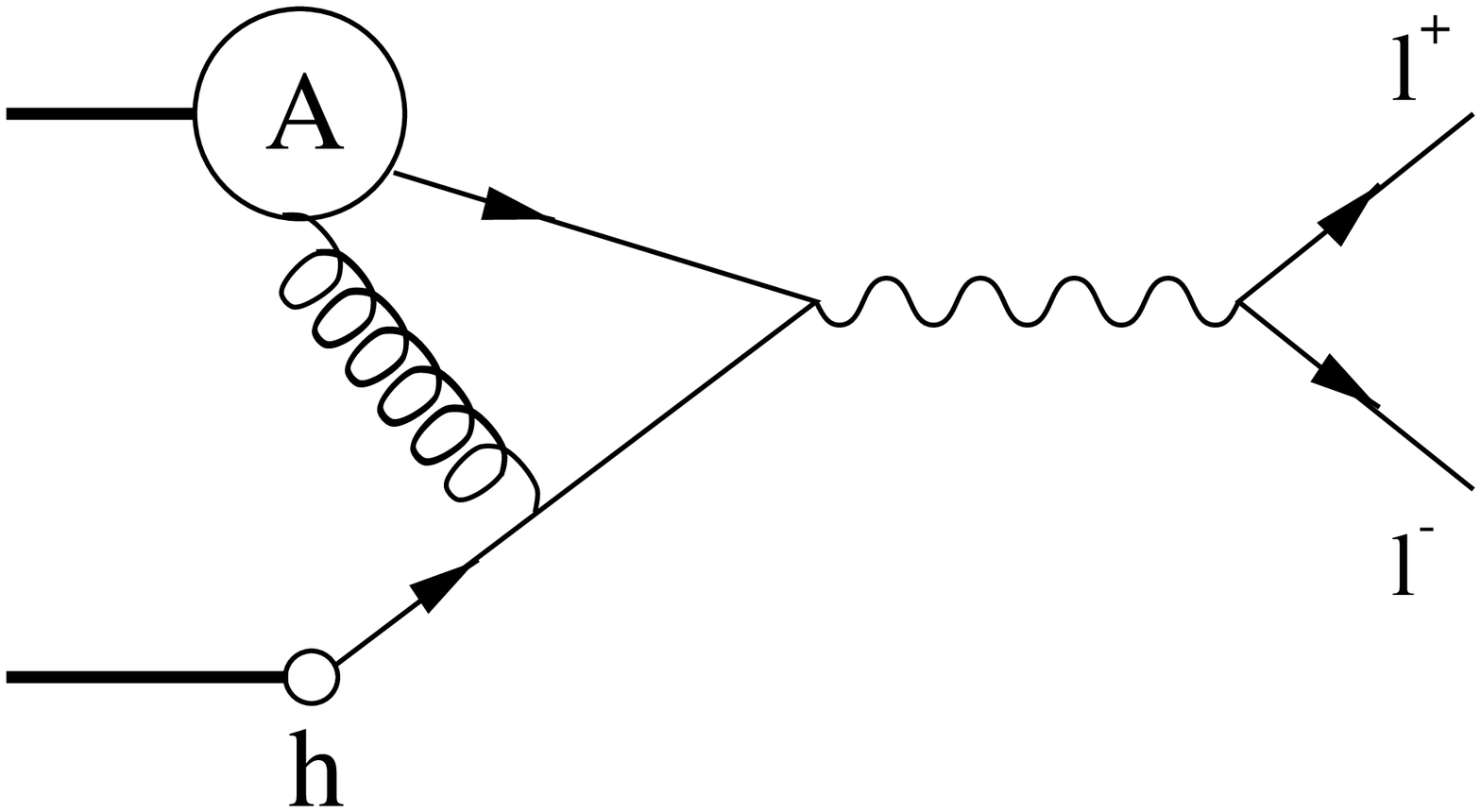} \hspace{3em}
   \includegraphics[height=2.8cm]{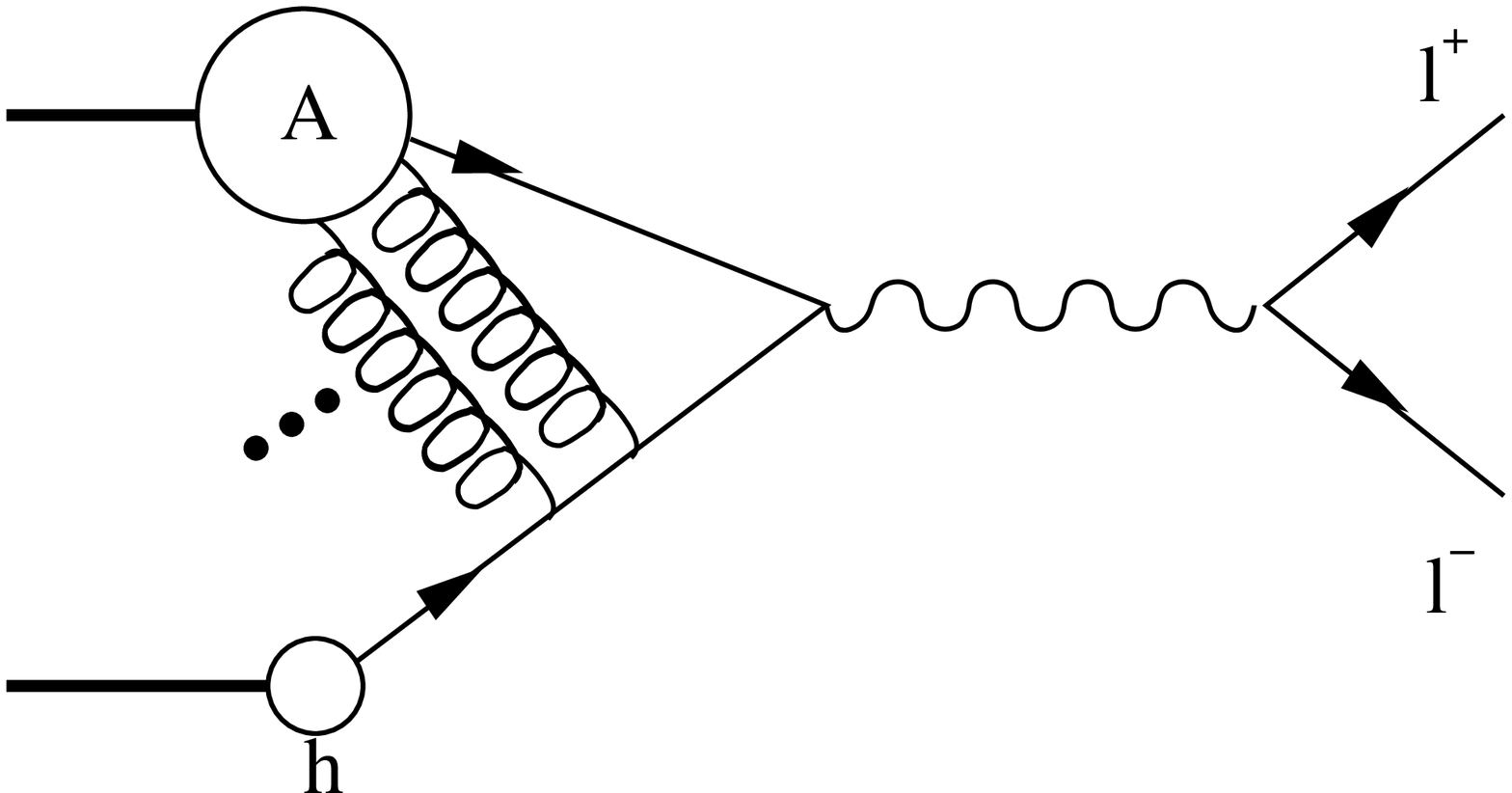}
   \\[0.5cm]
   \parbox{14cm}
        {\footnotesize 
        Fig.~6: DY graphs with scattering off one (left) and $n$ (right) 
        additional gluons from the nucleus.}
\end{center}

For the transverse momentum spectrum we obtain a shift and smearing.
The shift is given by $4\pi^2 \alpha_s A^{1/3} \lambda^2 /3$.
Some other topics addressed recently include the medium modification of
fragmentation functions \cite{GW:00} which can explain energy loss and the 
medium modification of parton distributions via higher twist modifications 
internal to the nucleus \cite{Qiu:LHC} which can account for shadowing.

Let us close with some remarks about the limitations of pQCD \cite{Qiu:LHC}. It
is well known that QCD factorization breaks down in Drell Yan beyond the order 
of twist-4 \cite{DFT}. Practically this means that there are non-factorizable 
contributions that make the twist expansion no longer reasonable beyond 
twist-4. However the nuclear enhancement argument saves us for $p+A$ 
collisions since it allows us to stay at twist-2 for the single proton. In 
that case we can go to arbitrary twist for the nucleus, like we can do also 
for $e+A$. For $A+A$ however the problem cannot be waived and we are limited
to twist-4 accuracy, which this still gives us the leading medium effect.


\vspace{-2mm}

\begin{thebibliography}{99}
\itemsep=0cm

\bibitem{CSS:mueller}
J.~C.~Collins, D.~E.~Soper and G.~Sterman
in A.~H.~Mueller (ed.): {\sf Pert.\ Quantum Chromod.},
World Scientific Publ., Singapore (1989).

\bibitem{LQS}
M.~Luo, J.~Qiu and G.~Sterman:
Phys.\ Lett.\ {\bf B 279}, 377 (1992); Phys.\ Rev.\ {\bf D 49}, 4493 (1994);
Phys.\ Rev.\ {\bf D50}, 1951 (1994).

\bibitem{Cron:75}
J.~W.~Cronin {\it et al.}:
Phys.\ Rev.\ {\bf D 11}, 3105 (1975).

\bibitem{E772:90dy}
D.~M.~Alde {\it et al.} [FNAL E772 Coll.]:
Phys.\ Rev.\ Lett.\  {\bf 64}, 2479 (1990).

\bibitem{McGMP:99}
P.~L.~McGaughey, J.~M.~Moss and J.~C.~Peng:
Ann.\ Rev.\ Nucl.\ Part.\ Sci.\ {\bf 49}, 217 (1999).

\bibitem{Guo:98ht}
X.~Guo:
Phys.\ Rev.\ {\bf D 58}, 036001 (1998).

\bibitem{FSSM:99}
R.~J.~Fries, B.~M\"uller, A.~Sch\"afer and E.~Stein:
Phys.\ Rev.\ Lett.\ {\bf 83}, 4261 (1999).

\bibitem{FSSM:00}
R.~J.~Fries, A.~Sch\"afer, E.~Stein and B.~M\"uller:
Nucl.\ Phys.\ {\bf B582}, 537 (2000).

\bibitem{LT:80}
C.~S.~Lam und W.~Tung:
Phys.\ Rev.\ {\bf D 21}, 2712 (1980).

\bibitem{Guo:98jb}
X.~Guo:
Phys.\ Rev.\ {\bf D 58}, 114033 (1998).

\bibitem{GW:00}
X.~Guo and X.~Wang:
Phys.\ Rev.\ Lett.\  {\bf 85}, 3591 (2000).

\bibitem{Qiu:LHC}
J.~Qiu: Talk at the Workshop {\sf Hard probes at LHC}, Geneva, Oct 2001.

\bibitem{DFT}
R.~Doria, J.~Frenkel and J.~C.~Taylor:
Nucl.\ Phys.\ {\bf B 168}, 93 (1980).


\end{thebibliography}
\end{document}